\documentclass{aa}

\usepackage{graphicx}

\newcommand{\teff}{\mbox{$T_\mathrm{eff}$}}
\newcommand{\logg}{\mbox{$\log g$}}
\newcommand{\halpha}{\mbox{H$\alpha$}}
\newcommand{\hbeta}{\mbox{H$\beta$}}

\newcommand{\uvby}{\mbox{$uvby$}}
\newcommand{\uvbyB}{\mbox{$uvby\beta$}}
\newcommand{\vsini}{\mbox{$v \sin i$}}

\newcommand{\ftot}{\mbox{$f_{\oplus}$}}
\newcommand{\Dmv}{\mbox{$\Delta m_\mathrm{V}$}}

\begin{document}

\title{On the anomaly of Balmer line profiles of A-type stars}
\subtitle{Fundamental binary systems
\thanks{partly based on DENIS data obtained at the European Southern
Observatory}
}

\author
{B. Smalley\inst{1},
R.B. Gardiner\inst{1} \thanks{\emph{Present Address:} The Meteorological Office,
Bracknell, UK},
F. Kupka\inst{2,3}
and M.S. Bessell\inst{4}}

\authorrunning{B. Smalley et al.}
\titlerunning{On the anomaly of Balmer line profiles}

\offprints{Barry Smalley \email{bs@astro.keele.ac.uk}}

\institute{
        Astrophysics Group, School of Chemistry \& Physics, Keele University,
        Staffordshire, ST5 5BG, United Kingdom\\
        \email{bs@astro.keele.ac.uk}
\and
        Institute for Astronomy, University of Vienna, T\"{u}rkenschanzstr. 17,
        A-1180 Vienna, Austria\\
        \email{kupka@astro.univie.ac.at}
\and
	Astronomy Unit, School of Mathematical Sciences, Queen Mary,
	University of London, Mile End Road, London, E1 4NS, United Kingdom\\
        \email{f.kupka@qmul.ac.uk}
\and
        Research School of Astronomy and Astrophysics, Institute of
        Advanced Studies, The Australian National University, Private Bag,
        Weston Creek P.O., ACT 2611, Australia\\
        \email{bessell@mso.anu.edu.au}
}

\date{Received <date> / accepted <date>}

\abstract{In previous work, Gardiner et al. (\cite{GKS99}) found evidence for a
discrepancy between the \teff\ obtained from Balmer lines with that from
photometry and fundamental values for A-type stars. An investigation
into this anomaly is presented using Balmer line profiles of stars in binary
system with fundamental values of both \teff\ and \logg. A revision of the
fundamental parameters for binary systems given by Smalley \& Dworetsky
(\cite{SD95}) is also presented. The \teff\ obtained by fitting \halpha\ and
\hbeta\ line profiles is compared to the fundamental values and those obtained
from \uvby\ photometry. We find that the discrepancy found by Gardiner et al.
(\cite{GKS99}) for stars in the range 7000~K $\la \teff \la$ 9000~K is no longer
evident.
\keywords{stars: general -- stars: atmospheres -- convection --
binaries: eclipsing -- stars: fundamental parameters} }

\maketitle

\section{Introduction}

Balmer lines are an important diagnostic of stellar atmospheric structure since
they are formed at a wide range of depths within the atmosphere. In
addition, the depth of formation of \halpha\ is higher than that of \hbeta,
thus observations of these profiles provide useful diagnostics (e.g. Gardiner
\cite{GAR00}). Balmer profiles are relatively insensitive to surface gravity
for stars cooler than $\sim$8000~K (Gray \cite{GRA92}; see also Heiter et al.
\cite{HEI+02}). In addition, Balmer profiles are sensitive to the treatment of
atmospheric convection (van't Veer-Menneret \& Megessier \cite{VM96};
Castelli et al. \cite{CGK97}; Gardiner \cite{GAR00}; Heiter et al.
\cite{HEI+02}). For stars hotter than $\sim$8000~K, the profiles are sensitive
to both effective temperature and surface gravity. However, provided we know
surface gravity from some other means (e.g. from eclipsing binary
systems), we can use them to determine effective temperature.

In previous work, Smalley \& Kupka (\cite{SK97}, hereafter SK97) found no
significant systematic problems with \uvby\ and fundamental (and standard)
stars. In fact, \uvby\ was found to be very good for obtaining \teff\ and
\logg. Using \halpha\ and \hbeta\ profiles, Gardiner et al. (\cite{GKS99},
hereafter GKS99) found that both the Canuto \& Mazzitelli (\cite{CM91,CM92}) and
standard Kurucz (\cite{KUR93}) mixing-length theory without overshooting
(MLTnoOV) (see Castelli et al. \cite{CGK97}) are both in agreement to within
the uncertainties of the fundamental stars. Overshooting models were always
clearly discrepant. However, GKS99 found some evidence for significant
disagreement between \emph{all} treatments of convection and fundamental values
around 8000 $\sim$ 9000~K. In this region the effects of \logg\ cannot be
ignored. In GKS99, most of the \teff\ stars do not have fundamental values of
\logg. Thus, possible \logg\ bias could have occurred. In this paper, we use
binary systems with fundamental values of \logg, determine revised fundamental
values of \teff\ and compare the results with those from Balmer lines.

Binary systems with fundamental values of \logg, with in some cases fundamental
values of \teff, were discussed by Smalley \& Dworetsky (\cite{SD95}, hereafter
SD95). Their list was limited to four systems, mainly due to lack of
trigonometric parallax measurements. Fortunately, the Hipparcos mission (ESA,
\cite{ESA97}) has provided trigonometric parallaxes for all of the binary
systems considered by SD95. Thus we re-evaluated 15 of those systems, using the
methods of SD95, with slight modifications and a discussion of the sources of
uncertainty.

Theoretical Balmer line profiles are compared to observations and the required
values of \teff\ are derived which when used in a model atmosphere will predict
the correct profiles. We have considered three models of convection: the
standard mixing-length theory {\sc atlas9} models (Kurucz \cite{KUR93};
Castelli et al. \cite{CGK97}), with and without approximate convective
overshooting, and modified {\sc atlas9} models based on the turbulent
convection theory proposed by Canuto \& Mazzitelli (\cite{CM91,CM92}) and
implemented by Kupka (\cite{KUP96}).

\section{Effective temperatures of binary systems}

Eclipsing binary systems provide ideal test stars for comparing to models,
since they enable us to obtain fundamental values of \teff\ and \logg. We can
obtain fundamental values of \teff\ provided we know the apparent angular
diameter ($\theta$) and total integrated (bolometric) flux at the Earth
(\ftot). In the case of binary systems, where there are no direct measurements
of angular diameters ($\theta$), we can obtain them from the stellar radius
($R$) and the parallax ($\pi$) of the system (for details see SD95). Since the
final \teff\ is twice as sensitive to $\theta$ than \ftot, it is critically
important to obtain good values of $\theta$, which requires both accurate
stellar radii measurements and accurate distance determinations.

Available spectrophotometry was taken from various sources (see
Table~\ref{flux_meas}), supplemented by 13-colour photometry fluxes from
Johnson \& Mitchell (\cite{JM+75}). All fluxes were placed on the Hayes \&
Latham (\cite{HL75}) absolute flux scale. Unfortunately, only four stars had
enough spectrophotometry. Many others, however, have at least $UBV$ or $UBVRI$
colours, which can be used to estimate the optical flux. Near-infrared fluxes
were taken from the Gezari et al. (\cite{GSM+87}), 2MASS (Skrutskie et al.
\cite{2MASS}) and DENIS (Epchtein et al. \cite{DENIS}) catalogues and placed on
the Mountain et al. (\cite{MOU+85}) absolute flux scale.

Using a method similar to Petford et al. (\cite{PET+88}), the total integrated
flux in the optical region can be obtained. Theoretical $UBV$ or $UBVRI$
colours were compared to the observed values. The model which gave best
agreement with the observations was then integrated from 3300--10000\AA.
While this method is not truly fundamental, the final integrated fluxes are not
that sensitive to the choice of models. An uncertainly of 10\%
was adopted to accommodate errors in the photometry and model fluxes.

In a few cases, not even $UBV$ colours are available. However, these systems
have Str\"{o}mgren $uvby$ photometry and hence ($B-V$) can be obtained from
($b-y$) using the relationship found by Crawford \& Barnes (\cite{CB70}).
Petford et al. (\cite{PET+88}) found that the ($B-V$) index could be used to
obtain the integrated flux from 3800--9000 \AA, to within about 3\% of that
from reticon fluxes. Hence, in the absence of other observations, the Petford
et al. empirical relationship is used, with an adopted uncertainty of 10\%, to
be consistent with that adopted for fluxes obtained from $UBV$ and $UBVRI$
photometry.

\begin{table}
\caption[ ]{Sources of flux measurements for binary systems.}
\label{flux_meas}
\begin{flushleft}
\begin{tabular}{lrl} \hline\hline
   \multicolumn{1}{c}{Star} & \multicolumn{1}{c}{HD} & Flux Sources \\ \hline
 \object{12 Per}       &  16\,739 & S2/68 B85 JM G87           \\
 \object{CD Tau}       &  34\,335 & S2/68 $UBV$ 2MA            \\
 \object{UX Men}       &  37\,513 & ($b-y$) 2MA                \\
 \object{$\beta$ Aur}  &  40\,183 & IUE J76 B85 K88 JM G87     \\
 \object{WW Aur}       &  46\,052 & S2/68 $UBVRI$ 2MA          \\
 \object{PV Pup}       &  62\,863 & ($b-y$) 2MA                \\
 \object{RS Cha}       &  75\,747 & $UBV$                      \\
 \object{HS Hya}       &  90\,242 & $UBV$ 2MA DEN              \\
 \object{RZ Cha}       &  93\,486 & ($b-y$) DEN                \\
 \object{$\gamma$ Vir} & 110\,379 & IUE M78 B85 G92 K88 JM G87 \\
 \object{DM Vir}       & 123\,423 & ($b-y$)                    \\
 \object{V624 Her}     & 161\,321 & IUE $UBVRI$ 2MA            \\
 \object{V1143 Cyg}    & 185\,912 & S2/68 $UBV$                \\
 \object{MY Cyg}       & 193\,637 & $UBV$                      \\
 \object{$\delta$ Equ} & 202\,275 & S2/68 B85 JM G87           \\
\hline
\end{tabular}
\begin{minipage}[h]{\columnwidth}{
Key:
IUE~~INES data from the IUE satellite,
S2/68~~Starlink STADAT data base (see Thompson et al. \cite{TOM+78}, Carnochan
\cite{CAR79}),
J76~~Jamar et al. (\cite{JAM+76}),
M78~~Macau-Hercot et al. (\cite{MAC+78}),
B85~~Burnashev (\cite{BUR85}),
G92~~Glushneva et al. (\cite{GLU+92}),
K88~~Kharitonov et al. (\cite{KHA+88}),
G87~~Gezari et al. (\cite{GSM+87}),
JM~~ Johnson \& Mitchell (\cite{JM+75}),
2MA~~ Skrutskie et al. (\cite{2MASS}),
DEN~~ Epchtein et al. (\cite{DENIS}).
}
\end{minipage}
\end{flushleft}
\end{table}

Reddening has been determined using {\sc uvbybeta\/} (Moon \cite{MOO85}) and
converted to $E(B-V)$ using the relationship $E(B-V) = 1.35 E(b-y)$ from
Crawford (\cite{CRA75}).  Where necessary, the observed fluxes were de-reddened
using the de-reddening routine from {\sc dipso\/} (Howarth et al.
\cite{HOW+97}).

Following SD95, having obtained the combined \ftot\ of the system, we can
determine the \ftot\ for each component by assuming that the difference in
\ftot\ between the two components is given by the observed magnitude
difference, \Dmv. This is a reasonable approximation for a system with nearly
identical components, but not necessarily valid for systems with markedly
different spectral types. In these cases, we must apply the bolometric
correction to obtain the difference in bolometric magnitudes in order to assign
the appropriate \ftot\ to each component. Bolometric corrections (BC) were
calculated using the polynomials given by Flower (\cite{FLO96}).
Table~\ref{fund_values} lists the values of fundamental radii and \logg\ values
and the observed \Dmv\, and gives the results of $\theta$, \ftot\  and \teff\
for each system. In most cases the values obtained using the Hipparcos
distances are in agreement with that obtained from other means (e.g. Infrared
Flux Method, $uvby$ photometry, etc.) and other determinations (e.g. Jordi et
al. \cite{JOR+97}, Ribas et al. \cite{RIB+98}). Three systems are anomalous.
Two of them, \object{PV Pup} and \object{DM Vir}, have large uncertainties in
their parallax measurements. The other system, \object{HS Hya}, appears to have
a very good Hipparcos parallax. Torres et al. (\cite{TOR+97}) argue that an
hitherto undetected third star in the system is likely to have affected the
trigonometric parallax. Hence, these systems will not be used in our analysis.

\begin{table*}
\caption{Fundamental values of \teff\ for binary stars. The ``combined''
\teff\ values are given in {\it italics}.}
\label{fund_values}
\scriptsize
\begin{flushleft}
\begin{tabular}{llllllllllll} \hline\hline
             & $\pi$            &                  &     &                   &                   & Sp.   & $\theta$          & \ftot               &             \\
Star         & ($10^{-3}$ \arcsec)            & \Dmv             &     & Radii             & \logg             & Types & ($10^{-3}$ \arcsec)             & (10$^{-6}$ W m$^2$) & \teff       \\ \hline

\object{HD\,16\,739}    & 40.52 $\pm$ 1.25 & 0.57 $\pm$ 0.04  &     & \multicolumn{2}{c}{BSF98} & & {\it 0.788 $\pm$ 0.034} & {\it 310.  $\pm$ 20.0} & {\it 6220 $\pm$ 168} \\
\object{12 Per}       &                  &                  &  A  & 1.58  $\pm$ 0.05  & 4.16  $\pm$ 0.03  & F8V   & 0.596 $\pm$ 0.026 & 195.  $\pm$ 12.8 & 6371 $\pm$ 176 \\
             &                  &                  &  B  & 1.37  $\pm$ 0.04  & 4.24  $\pm$ 0.03  & G2V   & 0.517 $\pm$ 0.022 & 115.  $\pm$ 5.06 & 6000 $\pm$ 143 \\[+1mm]

\object{HD\,34\,335}    & 13.66 $\pm$ 1.64 & 0.17 $\pm$ 0.14  &     & \multicolumn{2}{c}{RJT99} & & {\it 0.305 $\pm$ 0.037} & {\it 45.0  $\pm$ 5.00} & {\it 6177 $\pm$ 410} \\
\object{CD Tau}       &                  &                  &  A  & 1.798 $\pm$ 0.017 & 4.087 $\pm$ 0.010 & F6V   & 0.229 $\pm$ 0.028 & 24.3  $\pm$ 3.06 & 6110 $\pm$ 415 \\
             &                  &                  &  B  & 1.584 $\pm$ 0.020 & 4.174 $\pm$ 0.012 & F6V   & 0.201 $\pm$ 0.024 & 20.7  $\pm$ 1.61 & 6260 $\pm$ 397 \\[+1mm]

\object{HD\,37\,513}    & 9.93 $\pm$ 0.62  & 0.15 $\pm$ 0.10  &     & \multicolumn{2}{c}{AND91} & & {\it 0.171 $\pm$ 0.011} & {\it 14.0  $\pm$ 2.00} & {\it 6151 $\pm$ 293} \\
\object{UX Men}       &                  &                  &  A  & 1.347 $\pm$ 0.013 & 4.272 $\pm$ 0.009 & F8V   & 0.124 $\pm$ 0.008 & 7.48  $\pm$ 1.12 & 6171 $\pm$ 302 \\
             &                  &                  &  B  & 1.274 $\pm$ 0.013 & 4.306 $\pm$ 0.009 & F8V   & 0.118 $\pm$ 0.007 & 6.52  $\pm$ 0.54 & 6130 $\pm$ 233 \\[+1mm]

\object{HD\,40\,183}    & 39.72 $\pm$ 0.78 & 0.17 $\pm$ 0.14  &     & \multicolumn{2}{c}{NJ94} & & {\it 1.411 $\pm$ 0.032} & {\it 4500. $\pm$ 379.} & {\it 9077 $\pm$ 217} \\
\object{$\beta$ Aur}  &                  &                  &  A  & 2.769 $\pm$ 0.031 & 3.930 $\pm$ 0.010 & A1V   & 1.023 $\pm$ 0.023 & 2430. $\pm$ 250. & 9131 $\pm$ 257 \\
             &                  &                  &  B  & 2.627 $\pm$ 0.029 & 3.962 $\pm$ 0.010 & A1V   & 0.971 $\pm$ 0.022 & 2070. $\pm$ 139. & 9015 $\pm$ 182 \\[+1mm]

\object{HD\,46\,052}    & 11.86 $\pm$ 1.06 & 0.19 $\pm$ 0.23  &     & \multicolumn{2}{c}{AND91} & & {\it 0.294 $\pm$ 0.027} & {\it 108.  $\pm$ 12.0} & {\it 7827 $\pm$ 419} \\
\object{WW Aur}       &                  &                  &  A  & 1.883 $\pm$ 0.038 & 4.187 $\pm$ 0.019 & A5m   & 0.208 $\pm$ 0.019 & 58.7  $\pm$ 8.65 & 7993 $\pm$ 470 \\
             &                  &                  &  B  & 1.883 $\pm$ 0.038 & 4.143 $\pm$ 0.018 & A7m   & 0.208 $\pm$ 0.019 & 49.3  $\pm$ 5.01 & 7651 $\pm$ 401 \\[+1mm]

\object{HD\,62\,863}    & 8.10 $\pm$ 5.88  & 0.05 $\pm$ 0.27  &     & \multicolumn{2}{c}{AND91} & & {\it 0.162 $\pm$ 0.118} & {\it 43.0  $\pm$ 6.00} & {\it 8372 $\pm$ 3053} \\
\object{PV Pup}       &                  &                  &  A  & 1.542 $\pm$ 0.018 & 4.257 $\pm$ 0.010 & A8V   & 0.116 $\pm$ 0.084 & 22.0  $\pm$ 4.07 & 8361 $\pm$ 3060 \\
             &                  &                  &  B  & 1.499 $\pm$ 0.018 & 4.278 $\pm$ 0.011 & A8V   & 0.113 $\pm$ 0.082 & 21.0  $\pm$ 2.41 & 8383 $\pm$ 3053 \\[+1mm]

\object{HD\,75\,747}    & 10.23 $\pm$ 0.46 & 0.01 $\pm$ 0.16  &     & \multicolumn{2}{c}{AND91} & & {\it 0.302 $\pm$ 0.015} & {\it 88.0  $\pm$ 9.00} & {\it 7342 $\pm$ 266} \\
\object{RS Cha}       &                  &                  &  A  & 2.137 $\pm$ 0.055 & 4.047 $\pm$ 0.023 & A8V   & 0.203 $\pm$ 0.011 & 44.2  $\pm$ 5.56 & 7525 $\pm$ 307 \\
             &                  &                  &  B  & 2.338 $\pm$ 0.055 & 3.961 $\pm$ 0.021 & A8V   & 0.223 $\pm$ 0.011 & 43.8  $\pm$ 3.21 & 7178 $\pm$ 225 \\[+1mm]

\object{HD\,90\,242}    & 11.04 $\pm$ 0.88 & 0.17 $\pm$ 0.09  &     & \multicolumn{2}{c}{TOR97} & & {\it 0.181 $\pm$ 0.014} & {\it 14.0  $\pm$ 1.40} & {\it 5985 $\pm$ 282} \\
\object{HS Hya}       &                  &                  &  A  & 1.275 $\pm$ 0.007 & 4.326 $\pm$ 0.006 & F5V   & 0.131 $\pm$ 0.010 & 7.55  $\pm$ 0.80 & 6028 $\pm$ 290 \\
             &                  &                  &  B  & 1.216 $\pm$ 0.007 & 4.354 $\pm$ 0.006 & F5V   & 0.125 $\pm$ 0.010 & 6.45  $\pm$ 0.40 & 5936 $\pm$ 255 \\[+1mm]

\object{HD\,93\,486}    & 5.43 $\pm$ 0.63  & 0.00 $\pm$ 0.14  &     & \multicolumn{2}{c}{AND91} & & {\it 0.162 $\pm$ 0.019} & {\it 15.0  $\pm$ 2.00} & {\it 6440 $\pm$ 432} \\
\object{RZ Cha}       &                  &                  &  A  & 2.264 $\pm$ 0.017 & 3.909 $\pm$ 0.009 & F5V   & 0.114 $\pm$ 0.013 & 7.50  $\pm$ 1.11 & 6440 $\pm$ 444 \\
             &                  &                  &  B  & 2.264 $\pm$ 0.017 & 3.907 $\pm$ 0.010 & F5V   & 0.114 $\pm$ 0.013 & 7.50  $\pm$ 0.60 & 6440 $\pm$ 396 \\[+1mm]

\object{HD\,110\,379}   & 84.53 $\pm$ 1.18 & 0.00 $\pm$ 0.04  &     & \multicolumn{2}{c}{POP80} & & {\it 1.502 $\pm$ 0.179} & {\it 1960. $\pm$ 157.} & {\it 7143 $\pm$ 450} \\
\object{$\gamma$ Vir} &                  &                  &  A  & 1.35  $\pm$ 0.16  & 4.21  $\pm$ 0.017 & F0V   & 1.062 $\pm$ 0.127 & 978.  $\pm$ 80.5 & 7143 $\pm$ 451 \\
             &                  &                  &  B  & 1.35  $\pm$ 0.16  & 4.21  $\pm$ 0.017 & F0V   & 1.062 $\pm$ 0.127 & 978.  $\pm$ 41.3 & 7143 $\pm$ 433 \\[+1mm]

\object{HD\,123\,423}   & 2.91 $\pm$ 1.25  & 0.00 $\pm$ 0.14  &     & \multicolumn{2}{c}{LAT96} & & {\it 0.068 $\pm$ 0.029} & {\it 6.00  $\pm$ 1.00} & {\it 7928 $\pm$ 1735} \\
\object{DM Vir}       &                  &                  &  A  & 1.763 $\pm$ 0.017 & 4.108 $\pm$ 0.009 & F7V   & 0.048 $\pm$ 0.021 & 3.00  $\pm$ 0.53 & 7928 $\pm$ 1740 \\
             &                  &                  &  B  & 1.763 $\pm$ 0.017 & 4.106 $\pm$ 0.009 & F7V   & 0.048 $\pm$ 0.021 & 3.00  $\pm$ 0.28 & 7928 $\pm$ 1714 \\[+1mm]

\object{HD\,161\,321}   & 6.93 $\pm$ 0.74  & 0.79 $\pm$ 0.12  &     & \multicolumn{2}{c}{AND91} & & {\it 0.242 $\pm$ 0.026} & {\it 89.0  $\pm$ 9.00} & {\it 8222 $\pm$ 489} \\
\object{V624 Her}     &                  &                  &  A  & 3.030 $\pm$ 0.034 & 3.834 $\pm$ 0.010 & A3m   & 0.195 $\pm$ 0.021 & 60.0  $\pm$ 6.44 & 8288 $\pm$ 497 \\
             &                  &                  &  B  & 2.209 $\pm$ 0.034 & 4.024 $\pm$ 0.014 & A7V   & 0.142 $\pm$ 0.015 & 29.0  $\pm$ 2.55 & 8092 $\pm$ 474 \\[+1mm]

\object{HD\,185\,912}   & 25.12 $\pm$ 0.56 & 0.07 $\pm$ 0.11  &     & \multicolumn{2}{c}{AND91} & & {\it 0.441 $\pm$ 0.012} & {\it 110.  $\pm$ 11.0} & {\it 6418 $\pm$ 184} \\
\object{V1143 Cyg}    &                  &                  &  A  & 1.346 $\pm$ 0.023 & 4.323 $\pm$ 0.016 & F5V   & 0.315 $\pm$ 0.009 & 56.8  $\pm$ 6.32 & 6441 $\pm$ 201 \\
             &                  &                  &  B  & 1.323 $\pm$ 0.023 & 4.324 $\pm$ 0.016 & F5V   & 0.309 $\pm$ 0.009 & 53.2  $\pm$ 3.38 & 6393 $\pm$ 136 \\[+1mm]

\object{HD\,193\,637}   & 3.79 $\pm$ 0.87  & 0.03 $\pm$ 0.16  &     & \multicolumn{2}{c}{AND91} & & {\it 0.109 $\pm$ 0.025} & {\it 12.2  $\pm$ 1.22} & {\it 7434 $\pm$ 877} \\
\object{MY Cyg}       &                  &                  &  A  & 2.193 $\pm$ 0.050 & 4.008 $\pm$ 0.021 & F0m   & 0.077 $\pm$ 0.018 & 6.17  $\pm$ 0.76 & 7459 $\pm$ 891 \\
             &                  &                  &  B  & 2.193 $\pm$ 0.050 & 4.014 $\pm$ 0.021 & F0m   & 0.077 $\pm$ 0.018 & 6.00  $\pm$ 0.43 & 7408 $\pm$ 865 \\[+1mm]

\object{HD\,202\,275}   & 54.11 $\pm$ 0.85 & 0.00 $\pm$ 0.07  &     & \multicolumn{2}{c}{POP80} & & {\it 0.869 $\pm$ 0.025} & {\it 420.  $\pm$ 30.0} & {\it 6393 $\pm$ 147} \\
\object{$\delta$ Equ} &                  &                  &  A  & 1.22  $\pm$ 0.03  & 4.34  $\pm$ 0.02  & F7V   & 0.614 $\pm$ 0.018 & 210.  $\pm$ 16.5 & 6393 $\pm$ 156 \\
             &                  &                  &  B  & 1.22  $\pm$ 0.03  & 4.34  $\pm$ 0.02  & F7V   & 0.614 $\pm$ 0.018 & 210.  $\pm$ 8.90 & 6393 $\pm$ 115 \\
\hline
\end{tabular}
\begin{minipage}[h]{180mm}{
Key:
BSF98~~ Barlow et al. (\cite{BSF98});
RJT99~~ Ribas et al. (\cite{RJT99});
NJ94~~ Nordstr\"{o}m \& Johansen (\cite{NJ94});
AND91~~ Andersen (\cite{AND91});
TOR97~~ Torres et al. (\cite{TOR+97});
POP80~~ Popper (\cite{POP80});
LAT96~~ Latham et al. (\cite{LAT+96})
}
\end{minipage}
\end{flushleft}
\end{table*}

Several of the systems considered here have measured infrared colours which
enables us to employ the Infrared Flux Method (IRFM) developed by Blackwell \&
Shallis (\cite{BS77}) to determine the mean \teff\ of the binary system, which
can be compared to the ``combined'' \teff\ obtained using the fundamental
method. The results from the IRFM are presented in Table~\ref{irfm_values}. In
most cases there is good agreement with the fundamental values. However, for
three systems there is significant disagreement:

\begin{description}

\item[\bf \object{PV Pup}:] As noted above the fundamental value is unreliable,
but the IRFM value is also significantly hotter than that obtained from \uvbyB\
photometry (7140~K). Thus, we conclude that the IRFM result is unreliable,
possibly due to the uncertainty of the optical flux, which was obtained from
$b-y$ alone.

\item[\bf \object{$\gamma$ Vir}:] The IRFM result is cooler than the
fundamental value, but both agree to within the relatively large errorbar of
the fundamental value. Interestingly, SD95 obtained a value of \teff\ = 6750
$\pm$ 470~K using the van Altena et al. (\cite{van+91}) trigonometric parallax,
which is in excellent agreement with the IRFM.

\item[\bf \object{V624 Her}:] The IRFM is cooler than the fundamental value.
The fundamental value is, however, in agreement with Popper (\cite{POP84}) and
Ribas et al. (\cite{RIB+98}).

\end{description}

\begin{table}
\caption{Effective temperatures obtained from the Infrared Flux
Method (IRFM) for the binary systems. The values are the ``combined'' flux
\teff.}
\label{irfm_values}
\begin{flushleft}
\begin{tabular}{lll} \hline\hline
Star         & HD     & \teff          \\ \hline
\object{12 Per}       &  16\,739 & 6275 $\pm$ 229 \\
\object{CD Tau}       &  34\,335 & 5985 $\pm$ 203 \\
\object{UX Men}       &  37\,513 & 6204 $\pm$ 244 \\
\object{$\beta$ Aur}  &  40\,183 & 8858 $\pm$ 221 \\
\object{WW Aur}       &  46\,052 & 7795 $\pm$ 227 \\
\object{PV Pup}       &  62\,863 & 7759 $\pm$ 313 \\
\object{HS Hya}       &  90\,242 & 6256 $\pm$ 181 \\
\object{RZ Cha}       &  93\,486 & 6404 $\pm$ 260 \\
\object{$\gamma$ Vir} & 110\,379 & 6747 $\pm$ 189 \\
\object{V624 Her}     & 161\,321 & 7710 $\pm$ 218 \\
\object{$\delta$ Equ} & 202\,275 & 6293 $\pm$ 162 \\ \hline
\end{tabular}
\end{flushleft}
\end{table}

Having obtained fundamental \teff\ values for the binary systems, we now
present observations of the Balmer lines of these systems and the results of
fitting profiles using models with differing treatments of convection. We will
also refer to the IRFM values and values of \teff\ obtained from \uvby\
photometry.

\section{Observations}

The \halpha\ and \hbeta\ observations were made at the Observatorio del Roque
de los Muchachos, La Palma using the Richardson-Brealey Spectrograph on the
1.0m Jacobus Kapteyn Telescope (JKT) in 1997 October/November. A 2400
l\,mm$^{-1}$ holographic grating was used and a 1124 $\times$ 1124 pixel Tek
CCD, giving a resolution of 0.4{\AA} {\sc fwhm}. Further \halpha\ and \hbeta\
observations were made at the Mount Stromlo Observatory, Australia in February
2000 using the Cassegrain Spectrograph on the ANU 74 inch telescope. A 1200
l\,mm$^{-1}$ blazed grating was used,  giving a resolution of 0.35{\AA} {\sc
fwhm}. Additional \hbeta\ profiles were taken from the work of SD95.

The data reduction of the profiles taken in 1997 and 2000 was performed using
the Starlink {\sc echomop} software package (Mills et al. \cite{MWC97}). In
most cases the final spectra had a signal-to-noise ratio in excess of 100:1.
Instrumental sensitivity variations were removed from the \halpha\ profiles by
comparing to observations of stars with intrinsically narrow Balmer profiles,
for example early-B or O type stars and G type stars, and the \hbeta\ profiles
were normalized such that the observed profile of Vega agreed to a model with
\teff=9550~K, \logg=3.95, [M/H]=$-$0.5 (Castelli \& Kurucz \cite{CK94}) and the
standard profiles of Peterson (\cite{PET69}).

\section{Effective temperatures from Balmer line profiles}

\begin{table*}
\caption{\teff\ obtained from Balmer line profiles for binary systems with known
values of \logg}
\label{balmer_results}
\scriptsize
\begin{tabular}{llllllllllllllllll} \hline\hline
             &   &                & \multicolumn{3}{l}{CM} && \multicolumn{2}{l}{MLT\_noOV 0.5} && \multicolumn{2}{l}{MLT\_noOV 1.25} && \multicolumn{2}{l}{MLT\_OV 0.5} && \multicolumn{2}{l}{MLT\_OV 1.25} \\
Star         &   & \teff\ (fund)  & \halpha & \hbeta & \uvby && \halpha & \hbeta && \halpha & \hbeta && \halpha & \hbeta && \halpha & \hbeta \\ \hline
\object{12 Per}       & A & 6371 $\pm$ 176 & 6250 & 6360 &      && 6350 & 6460 && 6300 & 6540 && 6500 & 6750 && 6550 & 6840 \\
             & B & 6000 $\pm$ 143 & 6030 & 6050 &      && 6030 & 6150 && 6030 & 6230 && 6030 & 6440 && 6030 & 6530 \\[+1mm]
\object{CD Tau}       & A & 6110 $\pm$ 415 & 6420 & 6250 & 6276 && 6300 & 6350 && 6400 & 6400 && 6600 & 6600 && 6650 & 6750 \\
             & B & 6260 $\pm$ 397 & 6420 & 6250 & 6311 && 6300 & 6350 && 6400 & 6400 && 6600 & 6600 && 6650 & 6750 \\[+1mm]
\object{UX Men}       & A & 6171 $\pm$ 302 & 6070 & 6170 & 6131 && 6000 & 6200 && 6050 & 6390 && 6095 & 6600 && 6140 & 6650 \\
             & B & 6130 $\pm$ 233 & 6030 & 6130 & 6094 && 5960 & 6140 && 6030 & 6340 && 6050 & 6550 && 6100 & 6630 \\[+1mm]
\object{$\beta$ Aur}  & A & 9131 $\pm$ 257 & 9350 & 9290 & 9586 && 9350 & 9290 && 9350 & 9290 && 9350 & 9290 && 9350 & 9290 \\
             & B & 9015 $\pm$ 182 & 9110 & 9160 & 9438 && 9110 & 9160 && 9110 & 9160 && 9110 & 9160 && 9110 & 9160 \\[+1mm]
\object{WW Aur}       & A & 7993 $\pm$ 470 & 8000 & 7950 &      && 8000 & 8050 && 8000 & 8250 && 8100 & 8250 && 8150 & 8450 \\
             & B & 7651 $\pm$ 401 & 7440 & 7490 &      && 7440 & 7590 && 7440 & 7790 && 7450 & 7770 && 7470 & 7990 \\[+1mm]
\object{RS Cha}       & A & 7525 $\pm$ 307 & 7520 & 7420 & 7739 && 7450 & 7500 && 7550 & 7700 && 7830 & 7620 && 7920 & 7930 \\
             & B & 7178 $\pm$ 225 & 7180 & 7080 & 7375 && 7095 & 7020 && 7180 & 7250 && 7540 & 7340 && 7540 & 7550 \\[+1mm]
\object{RZ Cha}       & A & 6440 $\pm$ 444 & 6240 & 6040 & 6278 && 6000 & 6050 && 6050 & 6250 && 6350 & 6360 && 6400 & 6500 \\
             & B & 6440 $\pm$ 396 & 6240 & 6040 & 6357 && 6000 & 6050 && 6050 & 6250 && 6350 & 6360 && 6400 & 6500 \\[+1mm]
\object{$\gamma$ Vir} & A & 7143 $\pm$ 451 & 6940 & 6750 & 6906 && 6900 & 6750 && 6950 & 6950 && 7100 & 6950 && 7300 & 7300 \\
             & B & 7143 $\pm$ 433 & 6940 & 6750 & 6906 && 6900 & 6750 && 6950 & 6950 && 7100 & 6950 && 7300 & 7300 \\[+1mm]
\object{V624 Her}     & A & 8288 $\pm$ 476 & 7800 &      &      && 7750 &      && 7750 &      && 7920 &      && 8010 &      \\
             & B & 8092 $\pm$ 450 & 7770 &      &      && 7720 &      && 7720 &      && 7870 &      && 7990 &      \\[+1mm]
\object{V1143 Cyg}    & A & 6441 $\pm$ 201 & 6400 & 6360 & 6467 && 6340 & 6440 && 6440 & 6550 && 6530 & 6760 && 6580 & 6820 \\
             & B & 6393 $\pm$ 136 & 6370 & 6300 & 6467 && 6270 & 6440 && 6370 & 6500 && 6490 & 6750 && 6520 & 6780 \\[+1mm]
\object{MY Cyg}       & A & 7459 $\pm$ 891 & 7200 &      & 7079 && 7150 &      && 7200 &      && 7350 &      && 7600 &      \\
             & B & 7408 $\pm$ 865 & 7330 &      & 7223 && 7280 &      && 7330 &      && 7480 &      && 7580 &      \\[+1mm]
\object{$\delta$ Equ} & A & 6393 $\pm$ 156 & 6300 &      & 6203 && 6250 &      && 6300 &      && 6450 &      && 6600 &      \\
             & B & 6393 $\pm$ 115 & 6300 &      & 6203 && 6250 &      && 6300 &      && 6450 &      && 6600 &      \\
\hline
\end{tabular}
\end{table*}

The observed Balmer line profiles are fitted here to model spectra to compare
the derived \teff\ with that from fundamental methods. One of our main aims is
to investigate the performance of different models of convection. The following
convection models were used, using solar-metallicity Kurucz {\sc atlas} models:

\begin{enumerate}

\item Standard {\sc atlas9} (Kurucz \cite{KUR93}) models using mixing length
theory (MLT) without convective overshooting. The value of the MLT parameter
$\alpha$ is the standard value of 1.25. These will be referred to as
MLT\_noOV~1.25 models in this paper.

\item Standard {\sc atlas9} models using MLT without convective overshooting.
The value of the MLT parameter $\alpha$ is 0.5.  These will be referred to as
MLT\_noOV~0.5 models.

\item Standard {\sc atlas9} models using MLT with approximate overshooting. The
value of the MLT parameter $\alpha$ used is 1.25. These will be referred to as
MLT\_OV~1.25 models.

\item Standard {\sc atlas9} models using MLT with approximate convective
overshooting. The value of the MLT parameter $\alpha$ used is 0.5. These will
be referred to as MLT\_OV~0.5 models.

\item Modified {\sc atlas9} models using the Canuto \& Mazzitelli
(\cite{CM91,CM92}) model of turbulent convection.  These will be referred to as
the CM models.

\end{enumerate}

The synthetic spectra were calculated using {\sc uclsyn} (Smith \cite{SMI92};
Smalley et al. \cite{SSD01}) which includes Balmer line profiles calculated
using the Stark-broadening tables of Vidal et al. (\cite{VCS73}) and metal
absorption lines from the Kurucz \& Bell (\cite{KB95}) linelist. This routine
is based on the {\sc balmer} routine (Peterson \cite{PET69}; Kurucz
\cite{KUR93}). The spectra were rotationally broadened as necessary and
instrumental broadening was applied with {\sc fwhm} ${= 0.4}$ \AA\ to match the
resolution of the observations. The synthetic spectra were normalized at
${\pm100}$ \AA\  to match the observations. The values of  \teff\ were obtained
by fitting model profiles to the observations using the least-square
differences.  A microturbulence of 2 km\,s$^{-1}$ was assumed throughout for
both the model atmosphere line opacities and spectrum synthesis.

Two stars, \object{HD\,46\,052} and \object{HD\,161\,321}, are Am stars, with
[M/H]=0.23 and 0.29, respectively, calculated from the $\delta{m_0}$
calibrations, derived by Berthet (\cite{BER90}) and Smalley (\cite{SMA92}). For
these stars Balmer profile fits using the solar abundance grid model
atmospheres gave results for \teff\ which were around 50 $\sim$ 100~K hotter,
compared to models with the appropriate [M/H].

The Balmer line profiles become increasingly less sensitive to \logg\ below
$\sim$8000~K thus any errors in the values of \logg\ used will have no
significant effect on the results. Hotter than $\sim$8000~K the Balmer profiles
become progressively more sensitive to \logg, making it important to have a
fundamental value of \logg\ with a small error. For example, at 8000~K, a
change in assumed \logg\ of 0.5 dex would have the effect to changing the
temperature of the model profile an observation is fit to by $\sim$150~K.

\begin{figure*}
\centering
\includegraphics[height=23cm,clip]{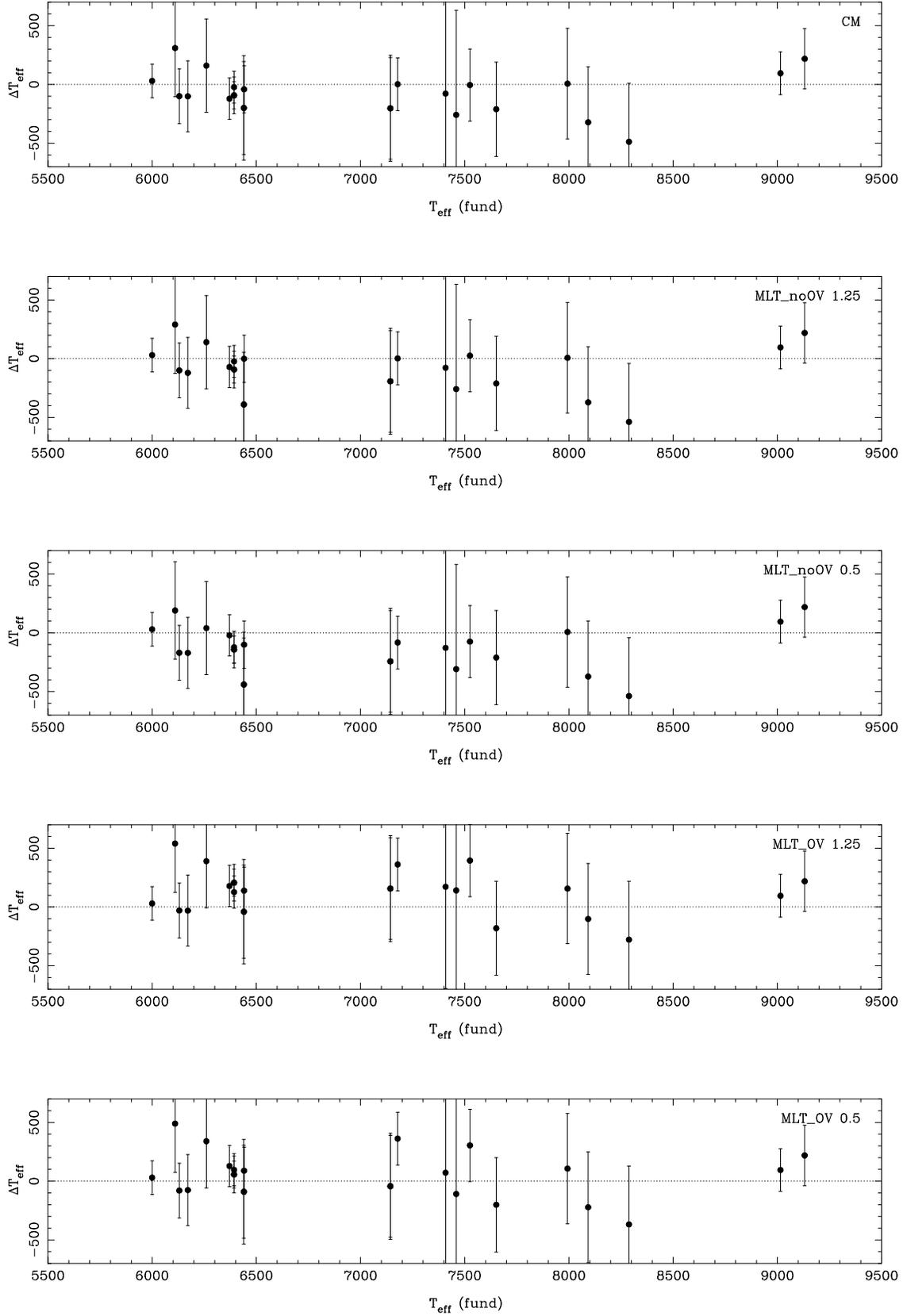}
\caption{Comparison between \teff\ calculated from Balmer line
profiles \halpha\ to those derived from fundamental methods for
the five convection models.
$\Delta$\teff\ $=$ \teff(Balmer) $-$ \teff(fund) is
plotted against \teff(fund). }
\label{halpha_comp}
\end{figure*}

\begin{figure*}
\centering
\includegraphics[height=23cm,clip]{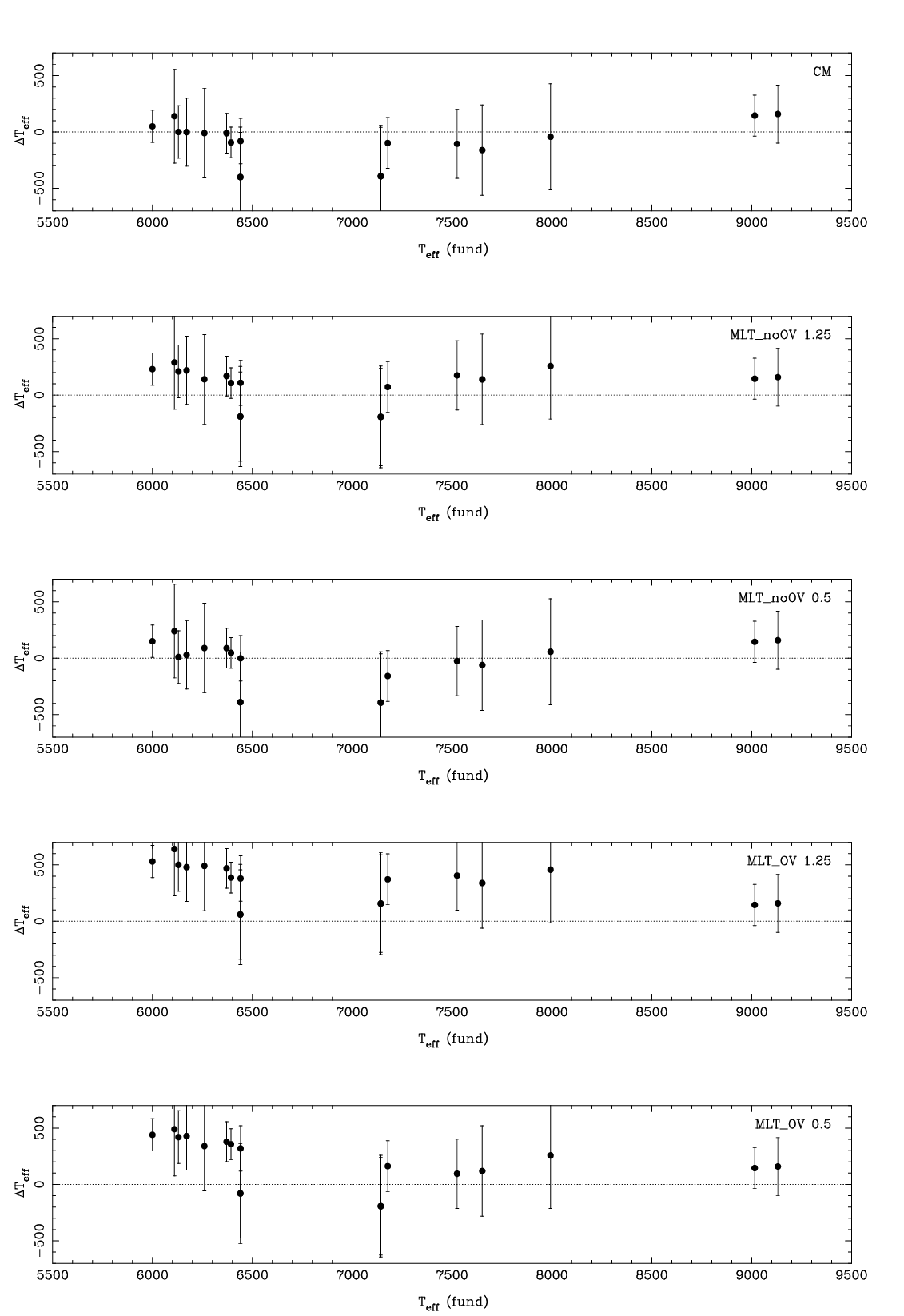}
\caption{as Fig.~\ref{halpha_comp}, but for \hbeta.}
\label{hbeta_comp}
\end{figure*}

The best fitting values of \teff\ for the binary stars are given in
Table~\ref{balmer_results}, for the 5 convection models listed above.
Figures~\ref{halpha_comp} \& \ref{hbeta_comp} show the variation of
$\Delta$\teff\ = \teff(Balmer) $-$ \teff(fund) against \teff(fund) for \halpha\
and \hbeta, respectively. To within the uncertainties, the CM results show no
significant variation with \teff(fund) for either \halpha\ or \hbeta. The
discrepancy around 8000~K noted by GKS99 is not evident. Even the two anomalous
\halpha\ points just hotter than 8000~K for \object{V624 Her} can be brought
into agreement if the IRFM \teff\ is used (see section 2). The MLT\_noOV
results are in broad agreement with those for CM, but with the $\alpha$=0.5
models giving better agreement around 8000~K relative to $\alpha$=1.25 and CM
models. Contrary to GKS99, who reported that F-type stars might require models
with $\alpha \ge$1.25 (see their Fig.~9), we find that the binary systems do
not support this. Overall, $\alpha$=0.5 models are preferred to those with
higher values. The MLT\_OV models are generally more discrepant, yielding too
high values of \teff\ (and even larger ones for \hbeta, if $\alpha$=1.25 is
used rather than 0.5), as found previously by GKS99. Note also the systematic
difference between \halpha\ and \hbeta\ for $\alpha$=1.25 MLT\_noOV models,
which is even more pronounced for the MLT\_OV models.

\subsection{The apparent A-star anomaly}

The use of stars with fundamental values of both \teff\ and \logg\ has failed to
support the apparent anomaly around 8000~K found by GKS99, in which Balmer
profiles gave progressively lower values of \teff\ compared to fundamental
values. However, there are too
few stars within the \teff\ range 8000--9000~K to fully explore this region. In
addition, GKS99 found that four fundamental \teff\ stars showed the
anomaly:
\object{HR\,4534}, \object{HR\,6556}, \object{HR\,7557}, and \object{HR\,8728}. In order to be sure that there is no
anomaly in the Balmer line profiles, we need to explain why these stars might
appear anomalous. We have also included \object{HR\,2421} which is just hotter than
9000~K, but find that this is in agreement with its fundamental and IRFM \teff.

\begin{table*}
\caption{Early A-stars with fundamental values of \teff, but not \logg.}
\scriptsize
\begin{tabular}{lllccccccccc} \hline\hline
HR   & star       & \vsini & \teff        & \logg & \multicolumn{2}{c}{CM \uvby} & \teff & \multicolumn{2}{c}{\halpha} & \multicolumn{2}{c}{\hbeta} \\
     &            & (km\,s$^{-2}$) & Fund       & GKS99 & \teff    & \logg & IRFM        & \teff & \logg & \teff & \logg \\ \hline
2421 & \object{$\gamma$ Gem} &  45  & 9220$\pm$330 & 3.5 & 9250$\pm$460 & 3.56$\pm$0.30 & 9040$\pm$86 & 9220$\pm$300 & 3.40$\pm$0.2 & 9060$\pm$250 & 3.52$\pm$0.2 \\
4534 & \object{$\beta$ Leo}  & 122  & 8870$\pm$350 & 4.1 & 8770$\pm$300 & 4.32$\pm$0.11 & 8660$\pm$60 & 8370$\pm$400 & 3.77$\pm$0.2 & 8450$\pm$350 & 4.07$\pm$0.2 \\
6556 & \object{$\alpha$ Oph} & 240  & 7960$\pm$330 & 3.8 & 7940$\pm$210 & 3.80$\pm$0.17 & 7883$\pm$63 & 7510$\pm$100 & 3.69$\pm$0.3 & 7580$\pm$150 & 3.42$\pm$0.6 \\
7557 & \object{$\alpha$ Aql} & 245  & 7990$\pm$210 & 4.2 & 7840$\pm$200 & 4.18$\pm$0.17 & 7588$\pm$73 & 7420$\pm$100 & 4.17$\pm$0.3 & 7450$\pm$150 & 4.38$\pm$0.6 \\
8728 & \object{$\alpha$ PsA} &  85  & 8760$\pm$310 & 4.2 & 8890$\pm$320 & 4.30$\pm$0.10 & 8622$\pm$86 & 8340$\pm$400 & 3.87$\pm$0.2 &      &      \\
\hline
\end{tabular}
\label{Astars}
\end{table*}

\begin{figure}
\centering
\includegraphics[width=\columnwidth,clip]{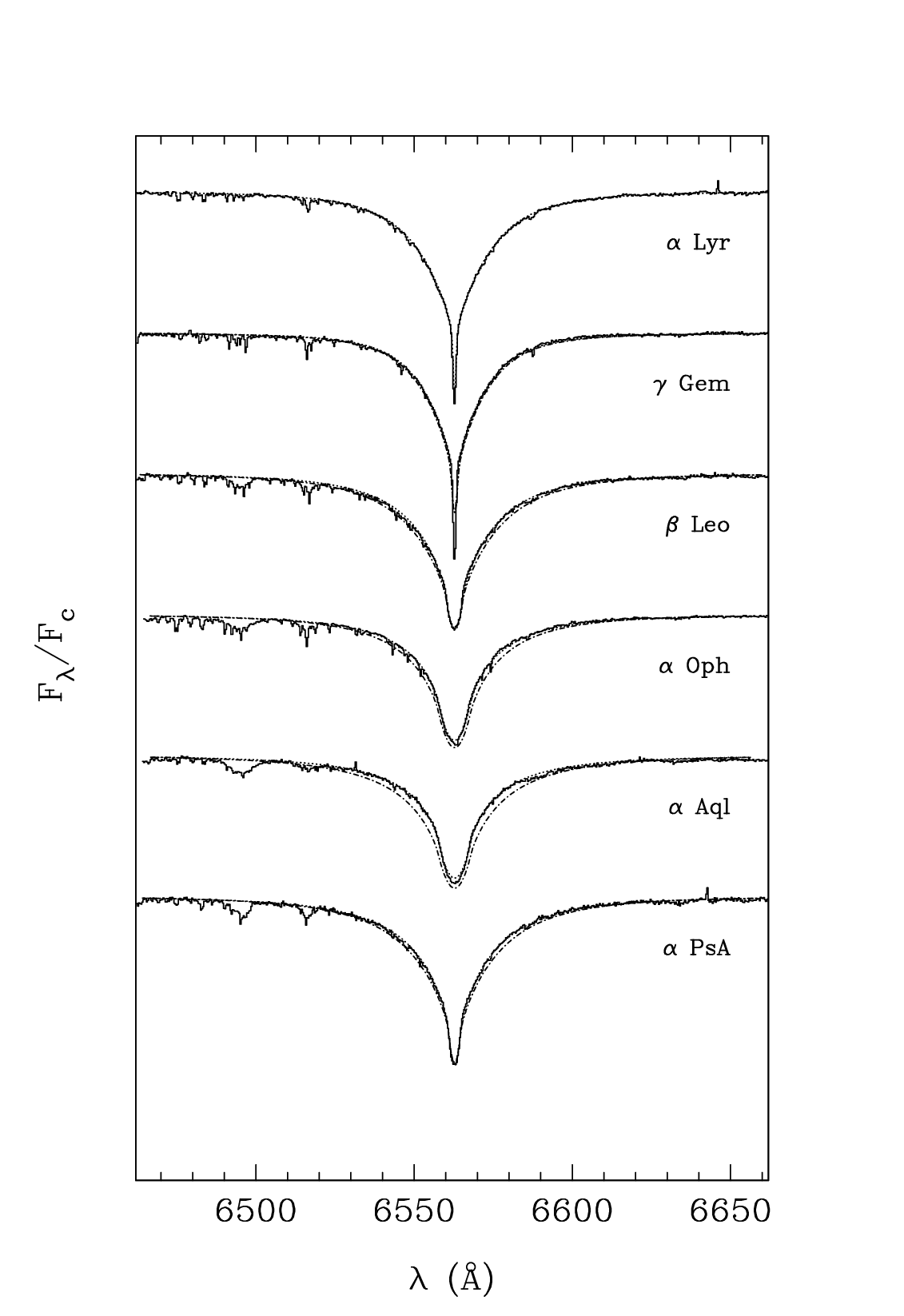}
\caption{\halpha\ profiles of A stars. The continuous line is the
observed profile, while the dotted line is the synthetic profile for
best fitting parameters given in Table~\ref{Astars}. The dash-dot
line is that for profiles calculated for the fundamental parameters.
The profiles for Vega are given for reference.}
\label{astars-halpha}
\end{figure}

\begin{figure}
\centering
\includegraphics[width=\columnwidth,clip]{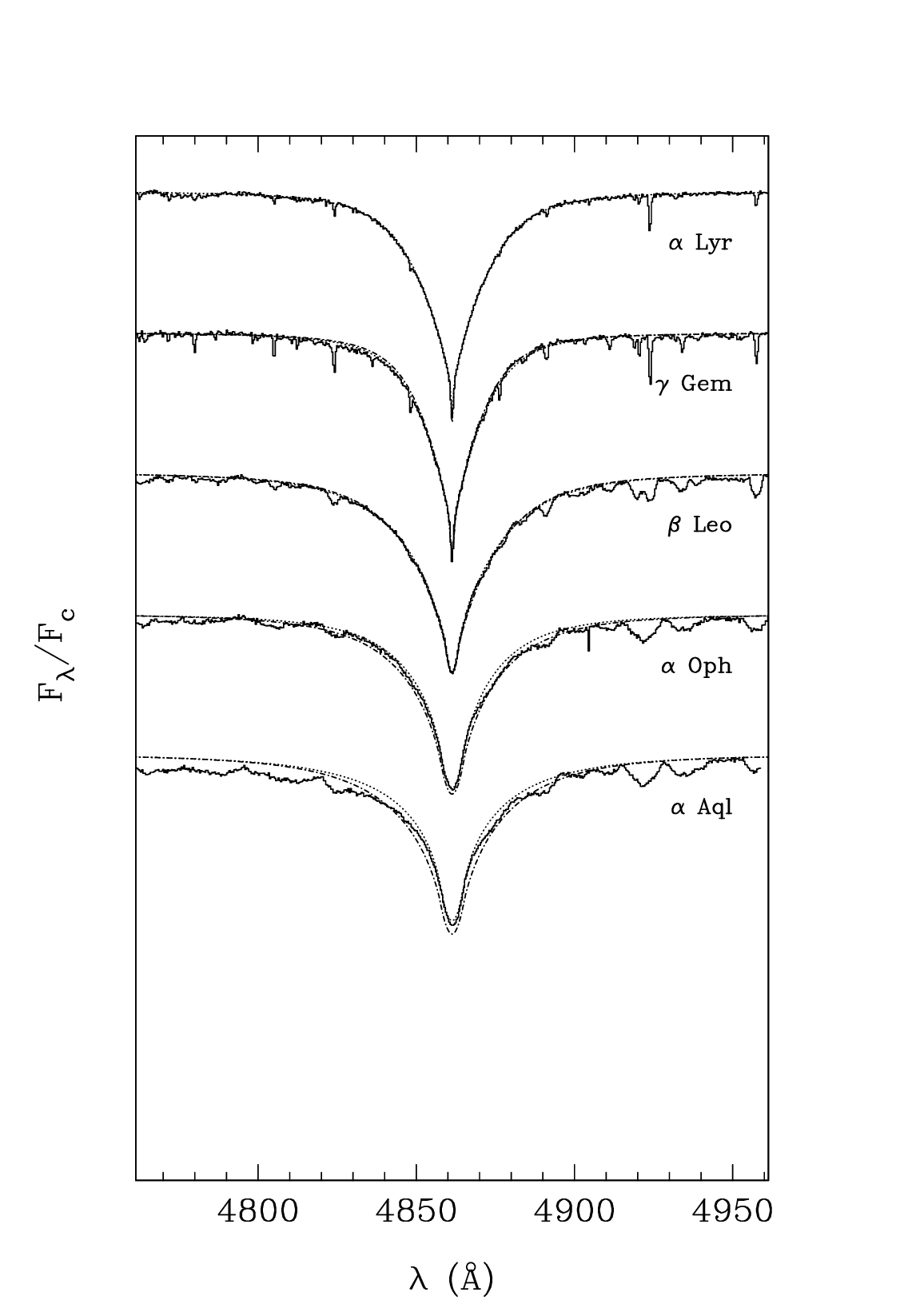}
\caption{as Fig.~\ref{astars-halpha}, but for \hbeta\ profiles. }
\label{astars-hbeta}
\end{figure}

Table~\ref{Astars} summarizes the values of \teff\ obtained from CM models and
\uvby\ photometry, the IRFM and by fitting to \halpha\ and \hbeta\ profiles. We
have allowed both \teff\ and \logg\ to vary in order to obtain the best
least-squares fit (see Figures~\ref{astars-halpha} \& \ref{astars-hbeta}).
Values of \logg\ are also given as obtained from \uvby\ photometry, as well as
those adopted by GKS99.

The rapidly rotating star \object{HR\,7557} has recently been studied by van
Belle et al. (\cite{vBEL+01}) using interferometry. Their analysis revealed the
oblateness of the star and gave revised angular diameter values and a new
determination of fundamental \teff\ = 7680$\pm$90~K. This is significantly
cooler than the previous determination, but in accord with that inferred from
the IRFM. As such, the \teff\ from \halpha\ and \hbeta\ are no longer
significantly discrepant. In fact the new \teff\ is in agreement with that
given in Gray (\cite{GRA92}) for a A7V star. It is certainly possible that
revision to the other fundamental stars could occur once new interferometric
measurements are obtained, especially \object{HR\,6556} which has a similar
\vsini\ and might be expected to exhibit significant oblateness. Thus, the
anomalies for \object{HR\,6556} and \object{HR\,7557} can be explained in terms
of their \emph{rapid rotation}. However, the earlier angular diameter
measurements are not necessarily the source of any discrepancy, since the new
\object{Vega} (\object{HR\,7001}) value obtained by Ciardi et al.
(\cite{CIA+01}) is in excellent agreement with that of Hanbury Brown et al.
(\cite{H-BDA74}).

The two other stars, \object{HR\,4534} and \object{HR\,8728}, have lower
\vsini\ values, but are the two most discrepant stars in the GKS99 sample.
Unless the fundamental values are truly wrong there must be some other reason
for the discrepancy. The IRFM values both point to a slightly cooler \teff, but
even then the discrepancy is $\sim$300~K. However, in this temperature region
the Balmer lines are near their maximum strength and sensitive to \logg. It is
certainly possible that a small error in adopted \logg\ could lead to a large
error in \teff\ obtained from Balmer profiles. In fact, using the values of
\logg\ obtained from $uvby$ photometry requires \teff\ $\simeq$7900~K in order
to fit the observed \halpha\ profiles for both stars and \teff\ $\simeq$8100~K
to fit the \hbeta\ profile of \object{HR\,4534}. In addition, the Balmer
profiles change little with relatively large changes in \teff. Thus, we
conclude that the two stars are not discrepant, due to the low sensitivity of
Balmer lines with respect to changes in \teff\ and both sensitivity to, and the
uncertainty in, the surface gravity for these stars.

In general, for stars hotter than 8000~K the sensitivity to \logg\ prevents us
from using Balmer line profiles to obtain values of \teff\ to the accuracy
required for the present task, unless we have accurate fundamental values of
\logg. However, until we do have stars with accurate fundamental \logg\ values,
we cannot be totally sure that there is not a problem with the model
predictions in this \teff\ region.

\section{Conclusion}

The availability of the Hipparcos parallax measurements has enabled the list of
stars with fundamental values of both \teff\ and \logg\ to be considerably
extended, from the 4 originally given by Smalley \& Dworetsky (\cite{SD95}), to
the 15 presented here. Even when the available optical flux measurements are
limited to only $UBV$ magnitudes, the quality of the final \teff\ values is
good. In some cases, it is the uncertainty of the Hipparcos parallax
measurements that limits the accuracy of the \teff\ obtained. The stars with
IRFM values are mostly in very good agreement with the fundamental values,
showing that the two methods are self-consistent and reliable. Since there are
more systems with \teff\ values from the IRFM (e.g. Blackwell \& Lynas-Gray
\cite{BL-G94}; Alonso et al. \cite{ALO+95}), this method can be used to obtain
`near-fundamental' values, provided we avoid binary systems with markedly
dissimilar components (Smalley \cite{SMA93}).

Balmer line profiles have been fitted to the fundamental binary systems. To
within the errors of the fundamental \teff\ values, neither the \halpha\ or
\hbeta\ profiles exhibit any significant discrepancies for the CM and MLT\_noOV
models. As in previous work, the MLT\_OV models are found to be discrepant.
Moreover, there are no systematic trends, such as offsets, between results from
\halpha\ and \hbeta\ as long as $\alpha$ in MLT models is chosen small enough
(e.g. 0.5). The discrepancies exhibited by the fundamental \teff\ stars in
GKS99 can be explained by rapid rotation in two cases and by the fact that the
Balmer profiles become sensitive to \logg\ and less sensitive to \teff\ in the
other two cases. However, for the time being the lack of any stars with
fundamental values of both \teff\ and \logg\ in this region precludes the
conclusion that there is not a problem with the models in the \teff\ range 8000
$\sim$ 9000~K.

\section*{Acknowledgments}

This work has made use of the {\sc simbad} data base, operated at the CDS,
Strasbourg, France, NASA's Astrophysics Data System (ADS) Abstract and Article
Service, and the INES (IUE Newly Extracted Spectra) archive. This work has also
made use of the hardware and software provided at Keele by the PPARC Starlink
Project. The referee, Mike Dworetsky, is thanked for his helpful comments on
the original manuscript. Rebecca Gardiner's PhD studentship was funded by
EPSRC. Friedrich Kupka acknowledges support by the project \emph{Turbulent
convection models for stars}, grant P13936-TEC of the Austrian Fonds zur
F\"orderung der wissen\-schaft\-lichen Forschung.

This publication makes use of data products from the Two Micron All Sky Survey,
which is a joint project of the University of Massachusetts and the Infrared
Processing and Analysis Center/California Institute of Technology, funded by
NASA and the NSF.

\end{document}